\documentstyle[preprint,revtex]{aps}                  
\begin{document}
\draft
\pagestyle{empty}                                     
\centerline{\hfill                 NTUTH--93--10} 
\centerline{\hfill                 August 1993} 
\vfill                                            
\begin{title}
Experimental and Theoretical Implications of \\
New Sequential Leptons
\end{title}
\vfill                                            
\author{Wei-Shu Hou and Gwo-Guang Wong}
\vfill                             
\begin{instit}
Department of Physics,
National Taiwan University,
Taipei, Taiwan 10764, R.O.C.
\end{instit}
\vfill                             
\begin{abstract}
If new sequential leptons $E^\pm$ and $N^0$ exist, the LEP bound implies
$m_E$, $m_N > M_Z/2$. The heaviness of the neutral lepton breaks away
from the pattern of the first three generations. The minimal model
is to have 4 left-handed lepton doublets and
4 right-handed charged lepton singlets, but only one
right-handed neutral lepton singlet.
Since in general the 3rd and 4th generation should mix,
and since $\vert m_N - m_E\vert$ should not be too large,
neither $E$ nor $N$ would be stable,
and both tend to decay via the Cabibbo suppressed $E\to \nu_\tau$ or
$N\to \tau$ charged currents.
This leads to the interesting signature of like-sign $W$ pair
production via $E^+N \to \bar\nu_\tau\tau^- W^+W^+$ at the SSC and LHC.
The popular seesaw mechanism cannot plausibly
accommodate the near masslessness of the light neutrinos
and the heaviness of $N^0$ simultaneously.
The representation structure poses a difficulty to
the traditional approach of $SO(10)$-based grand unified theories.
The discovery of such new heavy leptons would thus
have rather wide ranging and far reaching implications.
\end{abstract}

\vfill                             
\narrowtext
\def\ltap{\ \raisebox{-.5ex}{\rlap{$\sim$}} \raisebox{.4ex}{$<$}\ }
\def\gtap{\ \raisebox{-.5ex}{\rlap{$\sim$}} \raisebox{.4ex}{$>$}\ }
\def\vev#1{\left\langle #1\right\rangle}
\pagestyle{plain}

\section{Introduction}

Neutrino counting on the $Z^0$ resonance gives \cite{PDG} at present
\begin{equation}
   N_\nu = 2.99 \pm 0.04.
\end{equation}
Although the tau neutrino $\nu_\tau$ still needs to be
established as distinct from $\nu_e$ and $\nu_\mu$,
there is no denial that there exists three and only three
light neutrino species.
It is tempting to go one step further and state that there are
{\it only} three generations of fermions in Nature.
However, leaving prejudices aside, direct search for sequential leptons
at LEP yields the limits \cite{PDG},
\begin{equation}
m_E,\ m_N \gtap M_Z/2,
\label{eq:mN}
\end{equation}
where we denote the new sequential charged and neutral leptons as
$E^-$ and $N^0$, respectively. New Sequential leptons are permitted,
but they have to be very heavy, hence deviating sharply
from earlier generation patterns.

We in fact understand very little about fermion flavor.
Hence, the search for new sequential fermions must continue.
Although we have $m_\mu \sim m_s$ and
$m_\tau \sim m_b$, the neutrinos are {\it very} different
from up-type quarks in mass. The neutrinos are rather peculiar
in that they appear to be completely massless.
In the ``strict" Standard Model (SM), the left-handed neutrinos
are postulated to be massless by {\it assuming} the absence of
right-handed singlet partners.
Another popular way to achieve near vanishing neutrino masses,
motivated by grand unified theories (GUT) and the solar neutrino problem,
is the so-called seesaw mechanism \cite{seesaw}.
That is, left-handed neutrino
masses are driven to zero as $m_\ell^2/M_R$, where $m_\ell$
is the charged lepton (or $u$--type quark) mass,
by giving a very large
Majorana mass $M_R$ (usually of ${\cal O}(M_{\rm GUT})$)
to the corresponding right-handed $SU(2)\times U(1)$--singlet partner.

For up-type quarks, and in fact for all fermions that are known
to exist, the top quark is quite unique: It is very heavy.
We emphasize, however, that this  ``heaviness" is a matter of perspective.
The Yukawa couplings $\lambda_i$ break the chiral symmetry that arises
from the representation structure of Standard Model fermions.
 From the point of view of naturalness, the $\lambda$'s should all be
of order one if they are nonzero at all.
It is a mystery why these chiral symmetry breaking parameters
should be close to zero for the majority of known fermions.
In this respect, the top quark may be the only ``normal" fermion,
but if there were only 3 generations, it would be an endless
debate to ask whether it is the top being very heavy or all the
other fermions being too light.

If one or more new sequential generation exist,
the situation becomes more interesting.
The fourth generation $t^\prime$ quark is by definition heavier
than the top, while realistically speaking the $b^\prime$ quark shares
a similar hard limit as eq. (\ref{eq:mN}). Although the top is still unique,
we would then have a host of ``heavy" fermions where the
strengths of chiral symmetry breaking (Yukawa couplings) are similar
to those of the gauge couplings.
These would truly appear like ``normal" fermions at the
weak symmetry breaking scale $v.e.v. \simeq 246$ GeV,
with $\lambda_F \sim g_1$, $g_2 \sim 1$
($g_1$ and $g_2$ are electroweak gauge couplings),
while the ``light" fermions would be
grouped as (almost) chiral fermions with $\lambda_f \cong 0$.

Two important deviations from simple repetition are worthy of note.
The new heavy fermions should in general have
similar mass with respect to the top,
{\it i.e.} $\lambda_F \sim \lambda_t$, since
$\lambda_F \gg 1$ would be unreasonable even without considering
constraints like the $\rho$ parameter. This deviates from earlier
patterns in that generational hierarchies have to stop.
What we wish to emphasize and discuss further
in the present note, is the aforementioned deviation
for neutral leptons, that is $m_N \gtap M_Z/2$
(eq. (\ref{eq:mN})), which is in strong contrast to the
apparent $m_\nu \simeq 0$ for the first three generations.

A simple extension that could accommodate this
was recently pointed out by King \cite{King1}.
One adds a {\it single} right-handed neutral lepton singlet
$N_R$ to four sequential generations of the standard type.
In this way, if Majorana masses are forbidden,
one automatically has three strictly massless neutrinos.
We shall argue that, if new sequential leptons are found,
the extension of King would not only be minimal,
but would probably be the only plausible one. After working out the
parametrization of the model, we demonstrate that $E$, $N$
should have sizable mixings with $\tau$ and $\nu_\tau$,
and would thus undergo rapid decay.
The heaviness of $E$ and $N$ and the expectation that
$E$--$N$ splitting is not too large influence their decay properties,
leading to interesting implications for future search strategies.
We show that, when right-handed Majorana masses are added,
various modifications of the traditional seesaw mechanism
all have serious problems.
If the fourth generation has one extra $N_R$ singlet compared to
the first three generations, one would have problems grouping
the fermions into standard GUT multiplets.

\section{Parametrization of the Model}

The quark sector is composed of 4 standard generations,
hence we shall not comment further on it,
except recalling the fact that
$b^\prime$ may have unusual decay properties \cite{HS}.
Following King \cite{King1}, besides 4 left-handed lepton doublets
$\ell^\prime_{iL} = (\nu_{iL}^\prime$, $e_{iL}^\prime$)
and 4 right-handed charge $-1$ leptons $e_{iR}^\prime$,
we add just one right-handed (by convention) gauge singlet lepton $N_R$.
We shall call this situation ``$3+1$" generations.

We are interested in the couplings of fourth generation leptons.
Our approach is slightly different from that of refs. \cite{King1,King2}.
Assuming one Higgs boson doublet, upon symmetry breaking,
one has the lepton mass terms
\begin{equation}
\bar\nu^\prime_{iL}\, m_i\, N_R + \bar e_{iL}\, m_{ij}\, {e_j}_R,
\end{equation}
where we have assumed in addition that Majorana mass
for $N_R$ is forbidden (achieved, {\it e.g.}, by assigning some
unbroken, perhaps discrete, charge).
Note that $m_i$ can be chosen to be real.
The charged leptons can be diagonalised as usual by
a biunitary transform from the gauge basis $e^\prime_{L,\ R}$
to the mass basis $e_{L,\ R}$.
We define $e_{i} \equiv (e_{k},\ E)$ for both left and right-handed
components, where $e_k = e$, $\mu$, $\tau$ for $k = 1,\ 2,\ 3$.
The left-handed Dirac partner of $N_R$ is trivially chosen to be
\begin{equation}
\bar\nu^\prime_{iL}\, m_i \equiv \bar N_L m_N,
\label{eq:NL}
\end{equation}
where $m_N^2 = m_1^2 + m_2^2 + m_3^2 + m_4^2$,
and there is a single massive neutral lepton $N$ with Dirac mass
\begin{equation}
m_N \bar N_L N_R.
\end{equation}
The remaining 3 (left-handed) neutrinos remain strictly massless,
much like in SM. Because of this three-fold degeneracy,
one can arbitrarily redefine them without changing the physics.
We denote $\nu_{iL} \equiv (\nu_{kL}^0,\ N_L)$, $k = 1$--$3$.

The neutral current remains diagonal, and the standard $ZNN$ and $ZEE$
couplings lead to the bound of eq. (\ref{eq:mN}).
The leptonic charged current is (suppressing Dirac matrices)
\begin{equation}
\bar\nu^\prime_{iL} e^\prime_{iL}
   = \bar\nu_{iL}\, V_{ij}\, e_{jL},
\end{equation}
where
\begin{equation}
V^{\left(3+1\right)} = \left( \begin{array}{ll}
          \ U^{\left(3\right)}\ & \ 0\ \\
          \ 0\                  & \ 1\ \end{array} \right)
                K^{\left(3+1\right)}.
\label{eq:UK}
\end{equation}
In eq. (\ref{eq:UK}),
$U^{\left(3\right)}$ is a $3\times 3$ unitary matrix,
the zeros stand for three component column or row matrices,
and $K^{\left(3+1\right)}$ is the $4\times 4$ Kobayashi--Maskawa (KM)
fermion mixing matrix \cite{KM}.

Since the left-handed leptons are standard repetitions,
we have chosen to use the standard procedure to arrive first at
the KM matrix $K$, which possesses 6 rotation angles and
3 $CP$ violating phases.
The degeneracy of the three massless
neutrinos gives rise to the freedom of an arbitrary unitary matrix
$U^{\left(3\right)}$. We would like to see to what extent
the latter matrix reduces further the physical number of parameters
in $V$. One starts with $3^2 = 9$ parameters in the matrix $U$.
Three phases are absorbed by the unobservable phases of the
three massless neutrino fields. The 6 remaining parameters
consist of 3 angles and 3 phases. One immediately sees that
$V$ has {\it just three} mixing angles and no phases.
These physical angles describe the mixing
of $N_L$ with $e_L$, $\mu_L$ and $\tau_L$, respectively.

One can easily generalize
to $n+1$ generations (had there been $n$ light neutrino species!).
The $(n+1)\times (n+1)$ KM matrix $K$ would have
$n^2 = n(n+1)/2$ (angle)\ $+\ n(n-1)/2$ (phase) parameters,
while $U$ would have $n^2 - n = n(n-1)/2$ (angle) $+\ n(n-1)/2$ (phase)
parameters. All phase parameters are removed
and one is left with $n$ rotation angles,
precisely the number needed to describe the mixing between
the heavy neutral lepton and the $n$ light charged leptons.
Similarly, one can easily generalize to $3+m$ generations,
{\it e.g.} if $m$ new sequential generations exist and
one just adds $m$ new right-handed neutral lepton fields.
Following similar arguments, there should be $m(6+m-1)/2$
mixing angles, describing $3m$ angles between heavy and light
and $m(m-1)/2$ angles among the heavies, and
$(m-1)(6+m-2)/2$ phases, which can be similarly decomposed.
For the case of $m=2$, there should be $7$ angles and $3$ phases in
the lepton sector, compared to the KM prescription of 10 angles and 6 phases
in the quark sector. $CP$ violation effects may then occur in the lepton
sector in processes that involve the two new generations.
Further generalizations to $n+m$ generations is straightforward.

Returning to the $3+1$ case, it is useful to have an explicit
parametrization of the $3$ physical mixing angles.
We have to make a suitable choice of basis in the massless neutrino
sector. Recall that with 3 standard generations, {\it i.e.} when
$K$ is the $3\times 3$ KM matrix, $U$ is chosen such that
$V$ is the unit matrix. That is, the neutrinos are {\it defined}
to carry the label of the associated charged lepton, and
lepton number is separately conserved. Since the neutrinos
are physically degenerate, one takes the same unitary transform that
diagonalizes the charged lepton sector.
It is clearly advisable to stay close to this convention,
since the 3 mixing angles in the $3+1$ case just describes mixing
between $N_L$ and the three light charged letpons.
We therefore choose to build up $V$ by three such rotations \cite{HSS},
between $N_L$--$e_L$, $N_L$--$\mu_L$ and $N_L$--$\tau_L$,
where the rotations are denoted as $s_e,\ s_\mu$, and $s_\tau$,
respectively.
Thus, the lepton charged current is defined as
\begin{equation}
\left(\begin{array}[t]{rrrr}
\bar \nu_{eL}, & \bar\nu_{\mu L}, & \bar\nu_{\tau L}, & \bar N_L
      \end{array} \right)
       \left(
         \begin{array}{rrrr}
          c_e\ & -s_e s_\mu\ & -s_e c_\mu s_\tau\ \ & s_e c_\mu c_\tau \\
          0  \ &  c_\mu    \ & -s_\mu s_\tau    \ \ & s_\mu c_\tau     \\
          0  \ &  0        \ &  c_\tau          \ \ & s_\tau           \\
         -s_e\ & -c_e s_\mu\ & -c_e c_\mu s_\tau\ \ & c_e c_\mu c_\tau
         \end{array} \right)
        \left(
         \begin{array}{r}
          e_L \\
          \mu_L \\
          \tau_L \\
          E_L
         \end{array} \right).
\label{eq:CC}
\end{equation}

\section{Phenomenology}

The choice of zeros in eq. (\ref{eq:CC}) is quite arbitrary,
and we have adopted the convention that
$\bar\nu_{\mu L} e_L$, $\bar\nu_{\tau L} e_L$ and
$\bar\nu_{\tau L} \mu_L$ are absent.
Note, however, that lepton numbers are separately violated.
The physical observable (when massless neutrino states are involved)
is always the product $V_{ki}^T V_{kj}$ where $k$ is summed over
the 3 massless neutrino species.
For example, in the two neutrino experiment \cite{twonu},
the ratio of number of electrons produced versus muons
should be $s_e^2 c_e^2 s_\mu^2/(c_\mu^2 + s_e^2 s_\mu^2)^2$. The expected
smallness of $s_e$ and $s_\mu$, of course, makes this effectively unmeasurable.
Note that, if this experiment could be repeated at high energy,
the $\tau$ to $\mu$ ratio would be of order $s_\mu^2$.

We would like to explore the constraints on this
model, and, in particular, the expected properties of
the new leptons $E$ and $N$ \cite{King1,King2}.

\newpage

\subsection{Low Energy Constraints}

As remarked, clearly the $N_L$--$e_L$ and $N_L$--$\mu_L$ mixing
angles $s_e$ and $s_\mu$ should be rather small.
The best constraint is expected to be $\mu\to e\gamma$ and
$\mu \to e$ conversion on nuclei. The former gives \cite{AP}
$s_e^2 c_e^2 s_\mu^2 < 7\times 10^{-6}$, while the latter is expected
\cite{Marciano} to give the more stringent bound
\begin{equation}
s_e^2 c_e^2 s_\mu^2 < 10^{-8}.
\end{equation}
Although these are not separate bounds on $s_e$ and $s_\mu$,
they do suggest that $s_e$ and $s_\mu$
are extremely small,
and in any case these two angles are rather hard to test.

Based on observed patterns in the 3 generation quark sector,
the largest mixing angle is expected to be the $N_L$--$\tau$ angle
$s_\tau$. Although the ``$\tau$ decay puzzle" \cite{Marciano}
has largely evaporated with new $m_\tau$ measurements from
the BES Collaboration \cite{BES} and new $\tau$ lifetime measurements,
it is easy to see that within present error bars,
$N_L$--$\tau$ mixing can still be of Cabibbo strength. That is,
\begin{equation}
s_\tau \ltap 0.2
\label{eq:s34}
\end{equation}
is clearly permitted by present data.
 From the slightly more theoretical standpoint,
even if one assumes no
$N_L$--$\nu_\tau$ mixing ($\nu_{4L}^\prime \equiv N_L$ in eq. (\ref{eq:NL})),
the usual rule of thumb from quark mixing patterns leads to
$\tau_L$--$E_L$ mixing of order $\sqrt{m_\tau/m_E}$,
which ranges from $0.2 - 0.08$ for $M_Z/2 \ltap m_E < 300$ GeV,
quite similar to eq.  (\ref{eq:s34}).
Continued improvements on $\tau$ decay studies would remain the
best indirect searching ground for the existence of
new neutral leptons.

\subsection{Decay Properties of $E$ and $N$}

What is more exciting is, of course, the direct production
and detection of $E$ or $N$.
Most work in the past \cite{BHO} tend to
assume $m_N \ll m_E$, while some recent work has focused on
the case when $E$ and $N$ do not mix with light generations \cite{SY}.
The former is certainly no longer justified.
For the latter, we have seen that, although it is reasonable to
assume that $s_e$ and $s_\mu$ are vanishingly small,
$s_\tau$ can still be quite sizable.
We shall consider the charged current involving $E$ and $N$ as
effectively described by
\begin{equation}
\left(\begin{array}[t]{rr}
\bar\nu_{\tau L}, & \bar N_L
      \end{array} \right)
       \left(
         \begin{array}{rr}
          c_\tau\  & s_\tau           \\
         -s_\tau\  & c_\tau
         \end{array} \right)
        \left(
         \begin{array}{r}
          \tau_L \\
          E_L
         \end{array} \right).
\end{equation}

Since there is no reason to believe that $s_\tau$ is vanishingly
small in this model, whether $E$ or $N$ is the heavier one,
they would necessarily undergo rapid decay
because of their heaviness (eq. (\ref{eq:mN})).
Their decay rate is typically $s_\tau^2 \times \Gamma_t$ or higher,
where $\Gamma_t$ is the top quark decay rate
assuming $m_t \sim m_N$ or $m_E$,
even when $E$ and $N$ are degenerate.
Thus, the possibility of having stable charged or neutral leptons,
of relevance for cosmological considerations \cite{SY}
and for study at colliders \cite{FSY}, seems rather improbable.

Let us consider the case where $m_N > m_E \gtap M_Z/2$.
The $E$ would decay via $E\to \nu_\tau W^{\left(*\right)}$,
while there are two decay chains for $N$,
$N\to EW^{\left(*\right)}$ or $\tau W^{\left(*\right)}$.
Since $m_t > M_W$,
the $W$ boson, whether real or virtual,
would decay further via $W^{\left(*\right)}
\to e\bar\nu_e,\ \mu\bar\nu_\mu,\ \tau\bar\nu_\tau,\
\bar ud,\ \bar cs$. Ignoring $m_\tau$ and other ``light" fermion masses,
the decay rates are
\begin{eqnarray}
\Gamma(N\to EW^{\left(*\right)})\ &=&\ 9\times c_\tau^2\,
    \frac{G_F^2 m_N^5}{192\pi^3}\,
         f\left(\frac{m_N^2}{M_W^2},\frac{m_E^2}{m_N^2},
                                      \frac{\Gamma_W^2}{M_W^2}\right)
    \label{eq:NtoE},\\
\Gamma(N\to \tau W^{\left(*\right)})\ &=&\ 9\times s_\tau^2\,
    \frac{G_F^2 m_N^5}{192\pi^3}\,
         f\left(\frac{m_N^2}{M_W^2},0,\frac{\Gamma_W^2}{M_W^2}\right)
    \label{eq:Ntotau},
\end{eqnarray}
where the function $f(\rho,\mu,\gamma)$, accounting for decays
via both real and virtual $W$ bosons, can be found in \cite{BDKKZ}.
The $E\to \nu_\tau W^{\left(*\right)}$ rate is
identical to eq. (\ref{eq:Ntotau}) with $m_N$ replaced by $m_E$.
Na\"\i vely, one would have expected the ``Cabibbo favored"
$N\to E$ chain to be the dominant $N$ decay channel.
However, when $m_E$ is close to $m_N$, this chain
has rather limited phase space and may suffer from
$W$ propagator effects. There are also reasons such as $\rho$
parameter limits that suggest $N$--$E$ splitting should
not be too large, or else it would affect the global fit of
present day electroweak precision tests \cite{PDG}.
In contrast, the $N\to \tau$ sequence, though suffering
from ``Cabibbo suppression" through the factor of $s_\tau^2$ (in rate),
it does not suffer from phase space. Thus, it is not impossible
that the ``Cabibbo suppressed" $N\to \tau$ process
could in fact be dominant over the ``Cabibbo favored"
$N\to E$ process. This effect is displayed in Figs. 1 and 2.

With an eye towards the two major regions of experimental study
in the future, Figs. 1 and 2 are for the mass ranges
$m_{E,N} \in (50,\ 100)$ GeV and $(100,\ 300)$ GeV, respectively.
The dashed curves are for the $N\to \tau$ process.
The solid curves are for the $N\to E$ process for
$m_E = 50,\ 60,\ 70,\ 80,\ 90$ GeV for Fig. 1,
and $m_E = 100,\ 150,\ 200,\ 250$ GeV for Fig. 2.
For the latter set, we switch to dotdash lines for
$m_N > m_E + M_W$.
We illustrate with $s_\tau = 0.2$. With this $s_\tau$ value,
we see from Fig. 1 that $N\to \tau + W^*$ is typically
orders of magnitude higher than $N\to EW^*$. This would largely
hold even if $s_\tau$ is much smaller than 0.2, where one
can simply scale down the dashed curve.
For the heavier mass case, $N\to \tau$ also dominates over
$N\to E$ for the
plausible mass range $m_N - m_E \ltap M_W$, beyond which
the $N\to EW$ rate turns on sharply.
The eminence of $N\to \tau W^{\left(*\right)}$
has interesting implications on search strategies.

In case $m_E > m_N$, everything above holds true
upon making the interchange of $N\leftrightarrow E$ and
$\tau\leftrightarrow \nu_\tau$.

\subsection{Search Strategies}

In case $m_N$ or $m_E < M_W$,
one could search for $W\to \tau N$ or $E\nu$ \cite{King1}.
One could also indirectly check for sizable $s_\tau$
by studying $e$--$\mu$--$\tau$ universality since
$W\to\tau\nu$ would be suppressed by $1 - s_\tau^2$ \cite{King1}.
This test would demand rather high statistics and low
systematic background.

In the following, we focus on the
direct production of $E$ and/or $N$ pairs.
We shall always discuss the case $m_N > m_E$, as the opposite
can be easily reached by the interchange mentioned eariler.

The mass range $50$ GeV $< m_E,\ m_N < 100$ GeV would be of
immediate interest at LEP-II as soon as it turns on.
The $e^+e^-$ collider environment is rather clean such that
there should be no difficulty in finding $E^+E^-$ and $N\bar N$
pair production, although one suffers from low event rates.
The $E^+E^-$ pair results in the signature
$\nu_\tau\bar\nu_\tau{W^{\left(*\right)}}^+{W^{\left(*\right)}}^-$,
which is distinctive enough for $m_E$ below $m_W$.
Running at $\sqrt{s} < 160$ GeV would reveal the existence
of such charged leptons \cite{LEPII}.
However, for $m_E > M_W$, the signature becomes
$\nu_\tau\bar\nu_\tau W^+W^-$, and one is swamped by direct
$e^+e^- \to W^+W^-$ background that is
typically ten times larger \cite{LEPII}.
It is not clear whether the extra missing energy and the difference
in $WW$ angular distributions would be sufficient
to suppress background at LEP-II energies.

The purpose of Fig. 1 is to show that, with present knowledge that
$s_\tau$ could be as large as 0.2,
the neutral lepton $N$ would dominantly decay
via the Cabibbo suppressed
$N\to \tau W^{\left(*\right)}$ mode,
rather than the Cabibbo favored $N\to EW^*$ mode.
This is even more true when $m_N - m_E$ is small,
{\it i.e.} if $E$ is found at LEP-II first,
then the heavier it is, the lesser the likelihood that
$N$ would be discovered via the Cabibbo favored $N\to E$
channel, even with $s_\tau$ much smaller than $0.2$.
Hence, the discovery channel for $N$ is most likely
$N\bar N\to \tau^+\tau^-{W^{\left(*\right)}}^+{W^{\left(*\right)}}^-$.
Of course, if $m_E$ is close to $50$ GeV, and
$s_\tau$ is smaller than 0.2, it is possible to have
$N\to EW^*$ as the dominant $N$ decay mode.
In this case $N$ could be discovered via the decay sequences
$N\bar N\to E^-{W^*}^+ E^+{W^*}^-
\to \nu_\tau\bar\nu_\tau{W^{\left(*\right)}}^+{W^{\left(*\right)}}^-
{W^*}^+{W^*}^-$,
or
$N\bar N\to \tau^-{W^*}^+ E^+{W^{\left(*\right)}}^-
\to \tau^-\bar\nu_\tau {W^{\left(*\right)}}^+
{W^{\left(*\right)}}^-{W^*}^+$ ,
depending on the strength of $s_\tau$.

Note that, unless the mixing angle $s_\tau$ is
much smaller than $0.001 - 0.0001$,
both $N$ and $E$ should have sufficiently
short lifetime such that they would decay in the detector.
Of course, if $E^\pm$ is sufficiently long-lived, it could show up as
minimally ionizing charged tracks.
Note also that in case $m_E > m_N$ and $M_W$, if $s_\tau$ is sufficiently
small and $m_E - m_N$ is suitably large, the heavy $E$ may
be discovered via $E\to NW^* \to \tau {W^*}^+{W^*}^-$
{\it i.e.} in $e^+e^-\to \tau^+\tau^- {W^*}^+{W^*}^-{W^*}^+{W^*}^-$
above $W^+W^-$ threshold.

For $m_E,\ m_N > 100$ GeV, one would either need a high energy
$e^+e^-$ linear collider which should be able to cover the
full mass range $m_E,\ m_N \ltap \sqrt{s}/2$, or one would
have to resort to hadronic supercolliders such as the SSC or LHC.
It is usually claimed that heavy leptons that decay via
on--shell $W$ bosons suffer from large
vector boson pair production backgrounds and
would be difficult to detect at hadron supercolliders \cite{BHO,BP}
(for a counterview, see, however, ref. \cite{Hinch}).
However, in these earlier studies, it is usually assumed that the 4th
neutral lepton is massless. This is clearly no longer the case,
and it is of interest to see if the conclusions can be evaded.

The Drell--Yan production mechanism (via virtual $\gamma$, $Z$ or
$W$ bosons)
yields $E^+E^-$, $N\bar N$ and $E^-\bar N$ (or $E^+ N$) pairs,
with cross sections at the SSC ($\sqrt{s} = 40$ TeV)
ranging from $10 - 0.1\ pb$ as $m_{N}$ and $m_{E}$
range from $100 - 300$ GeV \cite{BHO}.
The background problem lies with $pp\to W^+W^- + X$ production, which is
of order $200\ pb$ and completely swamps the
$E^+E^- \to \nu_\tau\bar\nu_\tau W^+W^-$ process.
For the $N\bar N$ and $E\bar N$ modes, one needs to know
the $N\to E$ and $N\to \tau$ branching ratios.

It is seen from Fig. 2 that, with $s_\tau \simeq 0.2$ and
$m_N - m_E < M_W$, the Cabibbo suppressed $N\to\tau^- W^+$ mode
dominates (quite often by orders of magnitude) over
the Cabibbo favored $N\to E^-{W^*}^+$ mode.
Thus, for the $N\bar N$ production process,
the detection final state is $\tau^+\tau^-W^+W^-$.
Although the signal is one to three orders of magnitude
smaller than the $W^+W^-$ pair production background,
if the additional high $p_T$, isolated $\tau^+\tau^-$ pair can be utilized,
perhaps one could still separate the signal.
Detection in the $\tau^+\tau^- + 4\ jets$ mode may in fact
allow reconstruction of $m_N$ using kinematic tricks \cite{tautrick}.

What is more exciting is the $E\bar N$ mode. In our scenario
that $N\to \tau$ decay is likely to dominate over $N\to E$, we find
that $E^+ N$ decays into the final state
$\bar\nu_\tau \tau^- W^+W^+$. That is, the Drell--Yan production
of $E^+ N$ or $E^-\bar N$ leads to {\it like sign} $W$ boson pairs
plus isolated $\tau\nu$! The corresponding electroweak background
in this mass range is of order $1\ pb$ or less \cite{DV}.
Thus, the $\nu\tau^\mp W^\pm W^\pm$ signature should allow $E$ and
$N$ to be simultaneously discovered at the SSC (similar conclusions
should hold for the LHC). However, the additional neutrino
makes the reconstruction of $m_E$ and $m_N$ rather difficult.

As the $N\to EW^{\left(*\right)}$ branching ratio is raised,
which could come about if $s_\tau$ is considerably smaller than
0.2, or if $m_N - m_E > M_W$,
the signal switches to the more complicated
$E^+N \to
\nu_\tau\bar\nu_\tau W^+W^-{W^{\left(*\right)}}^+$,
and
$\bar NN\to
\nu_\tau\tau^\mp W^\pm W^\pm {W^{\left(*\right)}}^\mp$
or $\nu_\tau\bar\nu_\tau W^+ W^-
{W^{\left(*\right)}}^+{W^{\left(*\right)}}^-$.
That is, one may have triple or quadruple
(up to two being virtual) $W$ boson production
with additional associated handles like $\nu\bar\nu$ or
$\nu\tau^\pm$. In the corresponding case of $m_E > m_N$,
the final state $E^+ N$ or $E^-\bar N
\to \tau^+\tau^- W^+W^-{W^{\left(*\right)}}^+$
may allow for $m_N$ and $m_E$ reconstruction.

We conclude that the SSC and LHC should be able to
discover $N$ and $E$ with masses above $100$ GeV, especially
via $E^+N$ or $E^-\bar N\to \nu\tau^\mp W^\pm W^\pm$. However, further
detailed studies are needed to confirm this.

\section{Theoretical Implications}

It is foreseen that new limits on $E$ and $N$ would first come
from the onset of LEP II. This should allow the full exploration
of $m_N$, $m_E$ in the mass range up to half the beam energy,
of order 90 -- 100 GeV.
Beyond this, one would need either a high energy linear $e^+e^-$ collider,
or one would have to study the signatures discussed above at
the SSC or LHC. The Tevatron is not a good place for heavy lepton search.
It is of interest to ask,
if new sequential leptons are found, what would be the meaning
of such discoveries?
The implications turn out to be surprisingly profound
and wide ranging in scope, which should add
to the impetus and urgency for conducting heavy lepton searches.

\subsection{Problems with Seesaw Mechanism}

The most salient feature of discovering new sequential leptons
is the departure from previous patterns in first three generations:
The new neutral lepton must be rather heavy. This would pose
as a serious challenge to the usual seesaw mechanism for
explaining the near masslessness of known neutrinos.

In the standard seesaw mechanism \cite{seesaw}, one introduces
right-handed neutrinos for {\it each} neutrino species.
Since these extra fields are gauge singlets,
it is possible that they carry a lepton number violating
Majorana type of neutrino mass, denoted generically as $M_R$.
Assuming that the Dirac type of neutrino mass is
of order the corresponding charged lepton mass $m_{\ell^\pm}$,
if $M_R \gg m_\ell$, then the left-handed neutrino effectively
developes a Majorana mass of order $m_{\nu_L} \simeq m_\ell^2/M_R$.

Although $M_R$ can be quite arbitrary, and there are many tailor-made
models constructed for rather specific purposes \cite{Langacker},
the most popular and most natural setting for discussing
the seesaw mechanism is within GUT theories \cite{seesaw}, especially
GUT theories where right-handed neutrinos are incorporated
in multiplets together with other fermions. Not only the $M_R$ scale
gets independently motivated, it also seems \cite{Gallex}
to provide the best particle physics explanation for
the solar neutrino problem via the Mikheyev--Smirnov--Wolfenstein
(MSW) effect \cite{MSW}.

With three seemingly massless neutrinos, the seesaw mechanism
provides a rich playground for neutrino physics \cite{Langacker}.
However, if a new sequential neutral lepton $N$ is discovered,
one would have to reassess the utility of the mechanism.
Our (King) model is quite extreme in that we forbid
Majorana mass for $N_R$, and of course,
we have no right-handed neutrino fields for the first three
families. Thus, we cannot accommodate the seesaw mechanism.
Is it possible to construct models where the heaviness of $N$ and the
lightness of $\nu_k$, $k = 1$--$3$ are incorporated within
the framework of the seesaw mechanism?
Let us list the known options:
\begin{itemize}
\item[1)] Weak-scale seesaw: Motivated by dynamical symmetry
breaking ideas with $\bar t t$ condensation,
where one faces the problem of too heavy a top quark,
it was found desirable to introduce fourth generation fermions.
Hill and Paschos \cite{HP} proposed that $M_R$ is perhaps
of order 100 GeV. In this way, assuming $m_E$ is of similar order
of magnitude, small neutrino masses and eq. (\ref{eq:mN}) can
both be satisfied, but at a price. With the
seesaw mechanism retained, the model is rather precarious since
{\it all} the neutrino masses lie
{\it just} at the border of present limits \cite{HP}.
It is hard to believe that we are just at the juncture in
time such that the model is viable, although it certainly.
makes the model interesting in terms of immediate experimental
checks.
However, the main merit of the seesaw mechanism, the
MSW explanation for the solar neutrino problem, is lost,
since the 3 light neutrinos are too heavy.
\item[2)] Singular seesaw: Several groups \cite{Singular} have noticed that,
with more than one neutrino species, the right-handed neutrino
mass $M_R$ can be viewed as a $3\times 3$ matrix.
There is no strong reason to believe that this
matrix should be close to rank 3 in the sense that
$\det M_R$ could vanish, hence the name ``singular" seesaw.
Part of the original motivation was the 17 keV neutrino problem,
which has by now evaporated \cite{Hime}.
Applying the idea to the present case,
one envisions the standard type of seesaw for the first three
neutrinos, but for the fourth neutrino, $m_{4R}$
``accidentally" has a vanishing eigenvalue solution, hence the
resulting ``neutrino" is not necessarily very light.

There are at least two serious problems with this picture.
First of all, a high degree of tuning is needed to maintain
the smallness of $m_\nu$ for the first three generations and
satisfy eq. (\ref{eq:mN}) for the fourth ``neutrino".
This is especially so if one wants to invoke the MSW mechanism to
explain the solar neutrino problem, that is, when
$M_{kR} \sim M_{\mbox{GUT}}$ for $k = 1$--$3$, while
$m_{4R} \sim m_E$. This is reminiscent of the gauge hierarchy problem.
In the particular application to the fourth generation case,
Fukugita and Yanagida \cite{Singular} had to construct
a global $SU(4)$ family model that is broken in a complicated way
by multi-Higgs fields at some high scale.
Second, it seems artificial to have larger $M_R$ for
lighter generations while smaller for heavier ones.
\item[3)] Radiative seesaw: Babu and Ma \cite{BM} have constructed
a model similar to our $3+1$ model, but 
allowing for Majorana masses for $N_R$.
In this model,
Majorana masses are specifically given to $N_R$,
and one of the left-handed neutrinos acquires a seesaw mass
(that has to satisfy eq. (\ref{eq:mN})).
The chiral symmetry of the three originally massless left-handed
neutrinos is then broken, and they acquire radiative Majorana
masses through the two--loop two--$W$ graph.
The idea is interesting and may be explored further.
%
Although it may be difficult to conceive a realistic
working model, it can be viewed as
an extension of our $3+1$ model.
Again there is only one right-handed neutrino singlet.

\end{itemize}

\subsection{Problems with Grand Unification}

If the $3+1$ model is realized in Nature, it seems that,
as a corollary to the problem with the standard seesaw mechanism,
we would have to rethink our strategies regarding unifying
particle interactions.

The problem lies with the deviation from monotonous repetition
of representation structure. In most GUT theories,
the fermion generations are not unified, but rather, each generation
serves as one copy of the multiplet(s) structure of the GUT group.
The original GUT proposal, $SU(5)$ \cite{GG} puts
each generation into a $10$ plus a $\bar 5$, with the possible inclusion
of $\nu_R$ as a gauge singlet. This model is a direct generalization
of the Standard Model, and therefore could straightforwardly
accommodate $3+1$ generations. Having $SU(5)$ alone, however, is ruled
out by experiment.

Beyond $SU(5)$, it is customary to put each fermion
generation into one single multiplet, {\it e.g.} the
$16$ of $SO(10)$ has the $SU(5)$ decomposition of
$10\oplus \bar 5\oplus 1$, and therefore necessarily
requires a right--handed neutrino ($SU(3)\times SU(2)\times U(1)$ singlet)
for each generation. Originally \cite{seesaw}, this came hand-in-hand with
the seesaw mechanism and convinced many that right-handed neutrinos
exist, that the left-handed neutrino is extremely light, and one
had the extra bonus of providing a basis for explaining the
solar neutrino problem, as mentioned earlier.
Any GUT theory that puts each generation in one single multiplet
would face difficulty if generations repeat, but not entirely
sequentially, like in the $3+1$ case discussed here.

\section{Discussion and Conclusion}

The discovery of new sequential leptons
would provide great impetus for us to
reconsider traditional thinking in regards neutrinos, GUT theories,
and in particular, the question of fermion flavor.
The existence of a fourth generation with masses
$m_N,\ m_E,\ m_{b^\prime} \gtap M_Z/2$
and $m_{t^\prime} > m_t \gtap 91$ GeV
would give us a host of particles with Yukawa
couplings that are similar in order to the gauge couplings.
The pattern $m_1 \ll m_2 \ll m_3 \cdots$ has to
cease when $\lambda_F \sim m_F/v.e.v. \sim 1$ is approached.
The existence of 3 seemingly massless neutrinos while
the fourth neutral lepton is very heavy clearly breaks from
the traditional pattern.

It is clear that upon
discovery of such new particles, one has the burden to demonstrate
that one indeed has a new sequential family,
rather than, say, vector-like fermions where left- and right-handed
particles have the same representation structure.
That is, one would have to measure the gauge couplings
and demonstrate that $N$ and $E$ indeed form a left-handed
doublet, while $E_R$ and $N_R$ are weak singlets.
This is best done at an $e^+e^-$ machine, as LEP
has amply shown.

If such can be demonstrated, it is difficult to construct
a particle physics explanation for the solar neutrino problem.
If new experiments continue to suggest the existence of
such a problem, one may have to work along the lines of radiative
neutrino mass. In any case, the traditional, simple seesaw mechanism
cannot work.

In summary, we have explored the $3+1$ model of King,
where one adds just one right-handed neutral lepton singlet $N_R$ to
four generations of sequential fermions.
We elucidate the mixing properties
and demonstrate that there are only three rotational angles in
the lepton charged current, and no $CP$ violating phases. Such phases
start to appear if there are more heavy generations with associated
heavy neutral leptons. In the $3+1$ case, charged current mixing
is expected to be mostly in the $\nu_{\tau L}$--$N_L$ and
$\tau_L$--$E_L$ sector, described by one single mixing angle $s_\tau$
which could be as large as the Cabibbo angle.
Both $E$ and $N$ should be quite unstable,
even for the lighter of the two,
because of the possible decay chain $E\to \nu_\tau$ and $N\to \tau$.
The mass range $M_Z/2 < m_{N,E} < 100$ GeV can be explored soon
at LEP-II and also via $W$ decays.
If the mixing angle $s_\tau$ is not too small
and if $N$--$E$ splitting is not too large
($\vert m_N - m_E \vert \ltap M_W$), it is expected that
the Cabibbo suppressed decays $E\to \nu_\tau$ and $N\to \tau$
are the dominant ones for both $E$ and $N$,
whether $E$ or $N$ is lighter.
This leads to the distinctive like-sign $W$ boson pair
production signal $\nu\tau^\mp W^\pm W^\pm$ via $E^+N$ or $E^-\bar N$
Drell--Yan production at the SSC or LHC.
If they are found, the traditional seesaw mechanism,
and $SO(10)$ based GUT theories, would be at jeopardy,
and one may face a serious challenge with the solar neutrino problem.

\acknowledgments
WSH wishes to thank T. Han, C. Hill, C. S. Lim, E. Ma and E. Nardi for
discussions, and the Fermilab Theory Group for hospitality,
where part of this work is done.
This work is supported in part by grant NSC 82-0208-M-002-151
of the Republic of China.

\vskip -1cm
\figure{Decay rate for $N\to EW^{\left(*\right)}$ (solid)
and $\tau W^{\left(*\right)}$ (dashed) with
$s_\tau = 0.2$.
The solid curves
correspond to $m_E = 50$, $60$, $70$, $80$ and $90$ GeV.}
\vskip -1cm
\figure{Same as Fig. 2 except the solid curves
correspond to $m_E = 100$, $150$, $200$, $250$ GeV.
For $m_N > m_E + M_W$, the curves for $N\to EW^{\left(*\right)}$
switch to dotdash.}
\end{document}